\documentclass[useAMS,usenatbib]{mn2e}
\usepackage{graphicx}
\usepackage{aas_macros}
\usepackage{amsmath}
\usepackage{amssymb}
\newcommand{\z}[1]{$z\sim#1$} 
\newcommand{\R}{$\cal R$}
\newcommand{\Un}{$U_n$}
\newcommand{\U}{$U$}
\newcommand{\ciao}[1]{\textsc{#1}}   
\newcommand{\chandra}{\textit{Chandra}}

\newcommand{\lala}{LALA CETUS}
\newcommand{\lynx}{Lynx}
\newcommand{\egs}{EGS1}
\newcommand{\gws}{GWS}
\newcommand{\hdfn}{HDF-N}
\newcommand{\ergscm}{$\mathrm{erg\;s^{-1} cm^{-2}}$}
\newcommand{\ergs}{$\mathrm{erg\;s^{-1}}$}

\newcommand{\dd}{\mathrm{d}}
\begin{document}

\date{Accepted 2008 April 03. Received 2008 March 11; in original form 2007 November 02}

\pagerange{\pageref{firstpage}--\pageref{lastpage}} \pubyear{2002}

\title[X-ray luminosity function of AGN at $z\sim 3$]{The X-ray luminosity function of AGN at $z\sim 3$}
\author[J. Aird et al.]{J. Aird$^{1}$\thanks{E-mail:james.aird@imperial.ac.uk}, K. Nandra$^{1}$, A. Georgakakis$^{1}$, E. S. Laird$^{1}$, C. C. Steidel$^{2}$ and C. Sharon$^{2}$\\
$^{1}$Astrophysics Group, Imperial College London, Blackett Laboratory, Prince Consort Road, London SW7 2AZ, UK\\
$^{2}$California Institute of Technology, MS 105-24, Pasadena, CA 91125, USA}

\maketitle

\label{firstpage}

\begin{abstract}
We combine Lyman-break colour selection with ultradeep  ($\gtrsim 200$ ks) {\it Chandra} X-ray imaging over a survey area of $\sim 0.35$~deg$^{2}$ to select high redshift AGN.  Applying careful corrections for both the optical and X-ray selection functions, the data allow us to make the most accurate determination to date of the faint end of the X-ray luminosity function (XLF) at $z\sim3$. 
Our methodology recovers a number density of X-ray sources at this redshift which is at least as high as previous surveys, demonstrating that it is an effective way of selecting high $z$ AGN. 
Comparing to results at $z=1$, we find no evidence that the faint slope of the XLF flattens at high $z$, but we do find significant (factor $\sim 3.6$) negative evolution of the space density of low luminosity AGN.  Combining with bright end data from very wide surveys we also see marginal evidence for continued {\it positive} evolution of the characteristic break luminosity $L_{*}$. Our data therefore support models of luminosity-dependent density evolution between $z=1$ and $z=3$.
A sharp upturn in the the XLF is seen at the very lowest luminosities ($L_\mathrm{X} \lesssim 10^{42.5}$\ergs), most likely due to the contribution of pure X-ray starburst galaxies at very faint fluxes. 
\end{abstract}

\section{Introduction}
\label{sec:intro}
Understanding the evolution of active galactic nuclei (AGN), and thus the history of accretion in the universe, is an important problem in astrophysics. 
Accurate measurements to determine the shape of the luminosity function of AGN, and how this evolves with redshift, provide the key data to probe the changing distribution of AGN activity, and thus the importance of accretion onto super-massive black-holes at various times throughout the history of the universe.
Such investigations require large, unbiased samples of AGN spanning a range of redshifts and luminosities.

X-ray surveys are highly efficient at selecting AGN over a large range of redshifts \citep[e.g.][]{Barger03c, Eckart06, Brusa07}, including unobscured and moderately obscured AGN \citep[e.g.][]{Gilli07}, and very faint AGN where the light of the host galaxy overwhelms any optical signature \citep[e.g.][]{Moran02, Severgnini03}. Thus X-ray selected samples are relatively unbiased against all but the most heavily obscured objects.
Large efforts have been made to perform follow-up spectroscopic observations of X-ray sources detected in various surveys and measure their redshifts, allowing the evolution of the X-ray Luminosity Function (XLF) to be investigated.
AGN are found to be a strongly evolving population, with the number density decreasing dramatically from \z1 to the present day, similar to the star formation rate \citep[e.g.][]{Merloni04b}. Thus accretion activity, particularly in more luminous systems, was considerably more prevalent when the universe was about half its current age. 
This behaviour can be described by Pure Luminosity Evolution \citep[PLE,][]{Barger05}, in which the XLF retains the same shape (a double power-law with a break at a characteristic luminosity, $L*$), but shifts to higher luminosities as redshift increases. 
The most recent data indicate a complex picture, in which the redshift at which the AGN space density peaks is a function of luminosity \citep{Ueda03, Barger05, Hasinger05}. 
A number of authors have proposed that a Luminosity Dependent Density Evolution (LDDE) parameterization is necessary to describe such evolution \citep[e.g.][]{Miyaji00, Ueda03, Hasinger05, LaFranca05}.  In this scheme, the shape of the XLF changes with redshift, 
and various theoretical models have attempted to explain such evolution \citep[e.g.][]{Merloni04, Hopkins06b, Babic07}. 

The behaviour at higher redshifts remains unclear. While the most luminous AGN ($L_\mathrm{X}\gtrsim 10^{45}$\ergs) appear to continue to increase in number density at very high redshift  \citep[e.g.][]{Hasinger05, Silverman05}, very little is known about the behaviour of lower luminosity AGN. Very deep X-ray data are required to probe lower luminosities at these high redshifts, and thus there are few suitable fields and the sample sizes are small. In addition, the optical counterparts of faint X-ray sources are generally extremely faint, making spectroscopic identification difficult or sometimes impossible with current instrumentation. Even the most intensively--observed fields are therefore highly incomplete spectroscopically. In the \chandra\ Deep Fields -North and -South (CDF-N; CDF-S), the deepest \chandra\ observations to date with the most intensive programmes of spectroscopic follow-up, 87 per cent \citep[CDF-N:][]{Barger03c} and 78 per cent \citep[CDF-S:][]{Szokoly04} of X-ray sources with $R<24$ have spectroscopic redshifts. This completeness falls rapidly for objects with fainter optical counterparts, however, giving a total of 56 per cent and 39 per cent spectroscopically identified X-ray sources in the entire CDF-N and CDF-S samples respectively. This can severely bias determinations of the XLF, particularly at high redshifts \citep[e.g.][]{Barger05}. The key issue in determining the high redshift evolution is therefore how one deals with the sources which are too faint to identify. 

Two general approaches have been applied to address this. The first is to try to maximise the redshift completeness of the X-ray detected samples. Generally, this involves setting a relatively high X-ray flux limit, where spectroscopic completeness is higher, and augmenting the spectroscopy with photometric redshifts \citep[e.g.][]{Ueda03, Barger03c, Zheng04}. This allows redshifts to be assigned to most of the X-ray objects, and reduces the uncertainty in the determination of the high-redshift XLF \citep{Barger05, Hasinger05, Barger05b}. In the context of our present study, setting a high X-ray flux limit is highly undesirable, because the objects of interest (i.e. low luminosity AGN at high redshift) are at the limits of detectability even in the deepest X-ray surveys. In addition, there are considerable uncertainties in the redshift determinations, particularly for AGN. Finally, it is extremely difficult to correct accurately for any residual incompleteness in such samples, given that the success and failure rates of both spectroscopic and photometric redshifts, and errors in the photometric redshifts, depend on many complex factors. 

A completely different approach was taken by \citet{Nandra05b}. They used the Lyman-break surveys of \citet{Steidel03} to identify objects in a narrow redshift slice around \z3, based on their broad-band optical colours in three filters, with minimal contamination from lower redshift sources.
These were cross-correlated with X-ray detections in 2 deep X-ray fields to identify the presence of AGN. As the optical sample is limited to Lyman-break selected objects, lower significance detections of X-ray sources could be included, without increasing the chance of spurious alignments, thus probing to the maximum X-ray depth of  the available data. A sample of 10 AGN of moderate luminosity ($L_\mathrm{X}=10^{43-44.5}$\ergs) were identified, 9 of which have been spectroscopically confirmed. This colour pre-selection approach samples an incomplete, but well defined cosmological volume, which can be calculated from simulations \citep{Steidel99, Hunt04}. \citet{Nandra05b} presented the space density of moderate luminosity AGN at \z3, using the survey volumes given by \citet{Steidel99}. The result was significantly higher than previous work, and prompted the response of \citet{Barger05b}, using the most up-to-date multi-band optical and near infra-red photometric redshifts in the CDF-N to reduce incompleteness, increasing their result, bringing it to consistency with the completeness-corrected \citet{Nandra05b} value. In the present paper we extend and improve on this Lyman-break method of determining the high $z$ XLF, using three additional deep \chandra\ fields to increase our numbers of X-ray detected Lyman-break galaxies (LBGs), and improving the completeness corrections. 

In section \ref{sec:data} we describe the optical and X-ray data for our 5 fields, including the data reduction, selection of the Lyman-break sample and our X-ray source detection procedure. In section \ref{sec:selfunc} we describe our calculations of the X-ray selection function, and the optical selection function, tailored for our optical data and the selection of AGN counterparts. Section \ref{sec:lumfunc} describes our maximum-likelihood fitting to determine the XLF, carefully correcting for X-ray and optical incompleteness, and our improved method to determine binned estimates, accounting for the varying sensitivity of our observations. We compare our results to previous work in section \ref{sec:discuss} and discuss the implications for AGN activity at high redshift. 

A flat cosmology with $\Omega_{\Lambda}$=0.7 and $h=0.7$ is adopted throughout.

\section{Data/sample}
\label{sec:data}
The results presented in \citet{Nandra05b} were based on the 2 fields from the large Lyman-break survey of \citet{Steidel03} with sufficiently deep X-ray data: the \textit{Hubble} Deep Field-North (HDF-N), fully contained by the \chandra\ Deep Field-North with $\sim 2$ Ms exposure time, and the Groth-Westphal Strip (GWS) with $\sim 200$ ks exposure.
For this work we increase our sample using 3 additional fields with $\sim 200$ ks \chandra\ imaging with the ACIS-I instrument: the \lynx\ \citep{Stern02} and \lala\ \citep{Wang04b} archival fields, and \egs\ (part of the AEGIS-X survey, Nandra et al., in preparation), in addition to increasing our optical coverage of the \hdfn.
\subsection {Optical data}
\label{sec:optdata}
\subsubsection{Observations and data reduction}
\label{sec:optobs}
We have obtained deep optical imaging of our X-ray fields through \Un, G  and \R\ filters \citep{Steidel93}, suitable for Lyman-break selection at \z3.

Imaging of the \lynx\ and \lala\ fields was performed on the William Herschel Telescope (WHT) over 5 nights in November 2005, using the Prime Focus Imaging Camera which consists of a mosaic of two 2K $\times$ 4K EEV CCDs. The 16.2$\times$16.2 arcminute field of view is well matched to the \chandra\ ACIS-I field of view. The observations were performed as individual exposures of 500, 1000 and 1200 seconds in \Un, G and \R\ respectively, with the telescope being dithered by $\sim 30$ arcsec between exposures, to compensate for gain variations, bad pixels and the gap between the chips. \Un\ band data were always obtained closest to meridian to minimize the atmospheric extinction. The data were flat-fielded using twilight sky flats. For the \Un-band data additional flat-fielding was performed using dark sky `super-flats', produced by median combining the dithered exposures with objects masked out. Fringe removal was performed on the \R-band image using standard techniques. 

A number of spectrophotometric standard stars from the \citet{Oke90} catalogue were observed each night and were used to calculate photometric zero-points, corrected for Galactic extinction, as described in \citet{Steidel03}. The images were astrometrically calibrated by matching objects in the USNO-A2.0 astrometric catalog \citep{Monet98}, matching over 100 objects in each field and using a polynomial solution to map the focal plane to the catalogue positions. The solution was refined for each of the individual exposures to allow for small offsets and rotations. The individual exposures were then resampled to a common pixel scale. Final stacks in each band were produced by scaling each frame to match the zero-point and magnitudes of objects in the photometric exposure observed at the lowest air-mass. This allowed nonphotometric data to be included in the stacks, although exposures with poor seeing ($>1.5$ arcsec)  were excluded. Finally the stacked images were smoothed with a Gaussian filter so the the FWHM of stars matched in each photometric band (in practice this meant smoothing to match the \Un-band image quality). 

The \egs\ field was observed with the Palomar 200 inch telescope in May 2005, using the Large Format Camera, which consists of a mosaic of six SITe, back-side illuminated, 2048 $\times$ 4096 pixel CCDs. Individual exposures of 600 (\R), 1200 (G) and 1800 (\Un) seconds were obtained, with the telescope dithered inbetween. The data were flat-fielded using dome flats for the \R\ and G band images, and dark-sky flats for the \Un-band. Fringe removal was not necessary for the \R-band with the SITe CCDs. The astrometric and photometric calibration, and production of stacks was performed as for the WHT data. The \egs\ data are of the poorest quality in terms of depth and seeing in our sample, necessitating the inclusion of poorer seeing data in the stacks.

For the \hdfn\ field we use the higher quality, larger area imaging of GOODS-N region, used by \citet{Laird06}, rather than the original imaging presented by \citet{Steidel03} and used in \citet{Nandra05b}. Details of the observations and data reduction have been presented by \citet{Reddy06}. \textit{U}-band optical data were obtained with the KPNO Mosaic imager by the GOODS team, and transformed to \Un\ magnitudes. G and \R-band data were obtained on the Keck I telescope with the Low Resolution Imaging Spectrograph.

The optical data for the \gws\ used in this paper are identical to the `Westphal' field data presented by \citet{Steidel03}. The data were obtained on the Kitt Peak 4m Mayall telescope and the Palomar 200 inch in May 1996 and March 1997 respectively. The data reduction is fully described by \citet{Steidel03}, and is almost identical as our other fields.

The fields observed, telescopes used, the total exposure time (after excluding poor seeing data), and the average FWHM of stellar objects in the stacked images (before smoothing) are summarised in Table \ref{tab:fields}.

\subsubsection{Source detection, photometry and LBG sample selection}
\label{sec:optdet}
We perform 
source detection, deblending and photometry for all of our fields, using the  {\sc SExtractor} software  \citep{Bertin96}.
Objects were detected in the \R-band image, after smoothing with a 2 pixel FWHM Gaussian kernal, if they contained 5 or more connected pixels with flux exceeding $1.5\sigma_{\mathrm{RMS}}$ above the local sky background in the smoothed image, where $\sigma_{\mathrm{RMS}}$ is the pixel to pixel RMS noise in the unsmoothed image.
Isophotal apertures were defined using the detection criteria.  {\sc SExtractor} also attempts to deblend objects with merged isophotal apertures, but still resolvable as separate objects \citep[see][for details]{Bertin96}. 
The same isophotal apertures were then transferred to the images in other bands to measure isophotal magnitudes, which are used to calculate the
\Un-G and G-\R\ colours. However, the total magnitude we refer to when quoting \R\ magnitudes is measured within a flexible aperture which maximises the light collected, as defined by \citet{Kron80}.

A sample of \z3 LBGs is selected using the colour criteria given in \citet{Steidel03}:
\begin{eqnarray}
{\mathrm G}-{\cal R} &\le 1.2 \\
U_n- {\mathrm G} &\ge {\mathrm G} - {\cal R} +1.0
\end{eqnarray}
\label{eq:colsel}
Additionally objects are required to have 
\begin{equation*}
19.0\le {\cal R} \le  \left\{  \begin{array}{l l} 25.5    & {\mathrm{(Lynx,\ LALA\ CETUS)}}\\
							     25.0 & {\mathrm {(EGS1)}}\\
				\end{array} \right. 							     
\end{equation*}
and be detected with 3$\sigma$ significance or greater in the G band. The bright limit removes sources which may have contained saturated pixels. The faint limit ensures only high significance detections with accurate photometry are included in the LBG sample. We define 1$\sigma$ limits by $N_{\mathrm{pix}}^{0.5} \sigma_{\mathrm{RMS}}$ where $N_{\mathrm{pix}}$ is the number of pixels in the isophotal detection aperture.
There is no requirement on the \Un\ band magnitude, but objects with a
\Un\ band flux less than the 1$\sigma$ limit for their detection aperture are noted as
``un-detected'' and assigned a magnitude corresponding to the 1$\sigma$ limit.
The \egs\ data are of poorer quality and not as deep as the other fields. Thus, detection completeness is lower, and the photometry is less accurate, which results in more objects being scattered in or out of the LBG selection box, particularly at faint magnitudes. These effects will be accounted for in our calculations of the optical selection functions (section \ref{sec:optsel}). However, we increase the faint limit to \R$<25.0$ in \egs\ to prevent a high number of faint contaminants being scattered into our sample.

Our detection routine using  {\sc SExtractor} results in a different sample of LBGs to that presented by \citet{Steidel03} for \hdfn\ and \gws, using FOCAS to perform the detection and photometry. 
The overlap between our new samples and those previously presented is $\sim 50$ per cent in the \gws\ and $\sim 40$ per cent in the \hdfn (in the area covered by both data sets). A number of sources which were previously selected as LBGs no longer satisfy the colour criteria; many now satisfy the BX/BM criteria \citep{Adelberger04}, which selects objects in the region of colour space below the LBG selection, corresponding to objects at slightly lower redshifts. A number of additional objects are found in our LBG sample (in the same area covered by the original observations). Such scatter between different data sets and photometry is expected, particularly for fainter objects, and is accounted for in our calculation of optical selection functions (see section \ref{sec:selfunc}). While combining the previous and updated samples would increase our number of LBGs, the selection function would no longer be well-defined. The \hdfn\ and \gws\ LBG samples used for this work therefore consist of objects selected by our routines only.

In Table \ref{tab:fields} we give the total number of LBGs found in each field in the area covered by both X-ray and optical data.
%
%
\begin{table*}
\caption{
Fields used for this work. Column (1): field name; columns (2) and (3): Right Ascension (RA) and Declination (Dec) of the centre of the field; column (4): optical filter; column (5): telescope used and date for the optical observations; column (5): seeing FWHM of stellar objects in the field (prior to smoothing); column (6): total useful optical exposure of data with good seeing; column (7): Galactic column density as given by \citet{Dickey90}; column (8): X-ray exposure time after good time interval and background flare filtering; column (9): survey area with coverage by both X-ray and optical data; column (10) Number of Lyman-break galaxies detected in field, in the survey area.
}
\resizebox{\textwidth}{!} {
\begin{tabular}{c c c c c c c c c c c}
\hline
Field 		& RA		& Dec	& Filter 	& Telescope and Date	& Seeing 		& Optical &  $\mathrm{N_H}$	& X-ray 	& Survey & No. of 	\\
name		&		&		&		&					&			& exposure &				& exposure & area & LBGs	\\
			& (J2000) & (J2000) & 		& 					& (arcsec)   &	(s)						&  ($10^{20} \mathrm{cm}^{-2}$)& (ks) & (arcmin$^2$) & \\	
(1)			&	(2)	&     (3)	&   (4)       & 	(5)				&     (5)		 &    (6)	  &  (7)			& (8)		& (9)		& (10) \\
\hline
HDF-N		& 12:36:55.49 & +62:14:18.28 & \R & Keck I 2003 April  	&           0.7     	&    7200				& 1.5		& 1862.9	& 149.1	& 292 \\
			&			&			& G & 				& 	    0.9	&    7560					&		&		&		&	 \\
			&			&			& \U & KPNO 2002 April 	& 	    1.1	&    102600				&		&		&		&	\\
GWS			& 14:17:43.04 &  +52:28:25.20 & \R   & KPNO 1996 May/ 	&    	1.2		&	8300 			& 1.2		& 190.6    & 239.0 	& 326 \\
			&			&			& G	& P200 1997 Mar	&	1.2	&	7200 			& 		&		&		& 	\\
			&			&			& \Un &  				&	1.2		&	25200 	& 		&		&		& 	\\
Lynx			& 08:48:55.9 &  +44:54:50.0 	& \R   & WHT 2005 Nov	&    	1.0		&	5500 			& 2.7		& 186.5    & 243.4 	&  223 \\
			&			&			& G	& 				&	1.0		&	7000 			& 		&		&		& 	\\
			&			&			& \Un & 				&	1.1		&	12000 	& 		&		&		& 	\\
EGS1		& 14:22:42.66 & +53:25:24.83 & \R   & P200 2005 May	&    	1.2		&	3600 			& 1.2		& 177.8    & 358.9 	& 329 \\
			&			&			& G	& 				&	1.4		&	4500				& 		&		&		& 	\\
			&			&			& \Un & 				&	1.5  	&	14400 			& 		&		&		& 	\\
LALA 		& 02:04:44.25  &  -05:05:33.83 	& \R   & WHT 2005 Nov	&    	1.2	&	4500 			& 2.2		& 173.1    &  233.3	&  144  \\
CETUS		&			&			& G	& 				&	1.0		&	5000 			& 		&		&		& 	\\
			&			&			& \Un & 				&	1.3		&	22800			& 		&		&		& 	\\
\hline
\end{tabular}
}
\label{tab:fields}
\end{table*}

\subsection{X-ray data}
\label{sec:xraydata}
\subsubsection{Reduction}
The X-ray data for all the fields were reduced with our own pipeline procedure, which uses the \chandra\ Interactive Analysis of Observations (CIAO) software v3.1 and \chandra\ calibration database (CALDB) version 2.27. The reduction of the \gws\  and \hdfn\ fields is described in \citet{Nandra05} and \citet{Laird05} respectively. The same procedure was used to reduce the new fields, updating the charge transfer inefficiency corrections, applying the most recent gain maps, filtering out periods of high background, aligning the astrometry between different observations and producing images and exposure maps in our various standard bands (soft: 0.5-2 keV, full: 0.5-7 keV, hard: 2-7 keV, ultra-hard: 4-7 keV).
\subsubsection{Point Spread Function}
X-ray source detection requires knowledge of the Point Spread Function (PSF) of the \chandra\ telescope. We have calculated the PSF using MARX \citep{marxman}, simulating point sources with monochromatic energy 1 keV at a range of positions on the detector. For each position we determine the circular aperture which contains a particular Enclosed Energy Fraction. The PSF is defined at a position on the detector (fixed relative to the mirror). However, our deep X-ray data are all the result of a number of observations with different orientations and pointing directions, merged to create images with the maximum possible exposure. Thus, to calculate the PSF at positions in our images, we determined the EEF radius in each individual exposure. We then calculated the circular radius corresponding to the average area, weighting by the exposure per pixel in each observation. Further details of our PSF calculations will be given by Nandra et al. (in preparation).

\subsubsection{Initial source detection and astrometric correction}
\label{sec:xsrcdet}
X-ray source detection was initially performed on the soft-band image (0.5-2 keV) using the \ciao{wavdetect}\ tool in CIAO, with a low  threshold of $10^{-4}$. 
This is used to mask out sources when estimating the background in our actual source detection procedure (see below).
A culled version of the \ciao{wavdetect} catalogue, containing only higher significance detections (\ciao{src\_significance}$>4$), was also used to correct for astrometric offsets between the \chandra\ data and the optical imaging. This was done in two steps: first an initial cross-correlation of positions was performed using only secure, bright optical detections (\R$<23$) from the full  {\sc SExtractor} catalogue, with a search radius of 5 arcsec; a second matching was then carried out after correcting for any offset found, allowing fainter objects (\R$=23-24$) to be matched within a search radius of 2 arcsec.
This allows fainter counterparts to also be matched to the X-ray sources, increasing the number of matched objects and thus improving the accuracy of calculated astrometric offsets.

\subsection{X-ray counts extraction}
\label{sec:xopt}

Our approach to define a high-redshift sample of AGN is based on taking a well-defined, optically-selected sample of \z3 LBGs and determining those which host AGN. Thus, rather than searching for optical counterparts to an X-ray selected catalogue of objects, within a specified search radius 
we instead extracted soft-band counts at each LBG position to determine which LBGs have significant X-ray emission. Counts were extracted within circular apertures corresponding to the 70 per cent EEF for the PSF at that position.
The background level was estimated locally for each source from the number of counts within an annulus of inner radius equal to 1.5 times the 95 per cent EEF, and outer radius 100 pixels greater. All sources detected using \ciao{wavdetect} were masked out using 1.5 times their 95 per cent EEF radius for the background estimation.
The background count rate was then rescaled to match the effective exposure and area for the source extraction region. Finally we calculated the Poisson probability of the source region containing the observed counts or greater by chance, based on the background estimate. An LBG was `detected' in the X-ray data if this probability is less than $10^{-4}$. This relatively high threshold \citep[cf. $4\times 10^{-6}$ used by][]{Nandra05} may be used as we have a small number of test positions. Only $0.13$ false sources in total over all 5 fields are expected due to background fluctuations using this threshold.

A more problematic contamination, introducing false sources into our sample, is due to chance alignment of an LBG position with counts from a real X-ray source which are not associated with the actual LBG. There are 2 cases in which this can happen: i) counts from a nearby bright source are scattered into the LBG source region due to the PSF; ii) the LBG position falls on top of an X-ray source which is not otherwise associated with an optical counterpart.
Case i) is particularly a problem at high off-axis angles, where the PSF is large. To identify these contaminants, we extracted counts within the 95 per cent and 50 per cent EEF radii at each LBG position. 
When the LBG position matches the X-ray source position, the 95 per cent region should contain approximately $\frac{95}{50}=1.9$ times the number of counts in the 50 per cent region (above background, calculated as above). However, if the LBG is close to a bright source, and the detection is due to scattered counts, this factor will be larger, as the 95 per cent region will enclose a larger proportion of the bright source PSF distribution. Thus, we identify contaminants due to case i) when the 95 per cent region contains more than 4 times the number of net counts as the 50 per cent region. 13 significant detections (over all 5 fields) were identified as contaminants by this method, and were verified as such by visual inspection. 

It is not possible to distinguish between contaminants due to case ii) and when the LBG really is an X-ray source. However, we are able to estimate the expected number of false sources. To achieve this we generated random LBG positions by shifting the original positions by 30--60 arcsec in RA and Dec (this maintains any clustering), and repeated our source detection procedure.The process was repeated 100 times for each field. The average number of false detections in each random sample, after excluding contaminants due to case i) as above, were 1.01, 0.56, 0.46 0.8 and 0.41 in the \hdfn, \gws, \lynx, \egs\ and \lala\ fields respectively. The number varies between fields due to a number of factors, such as the depth of the optical data and therefore the density of LBGs, depth of the X-ray data, area coverage, and PSF effects. However, $\sim 70$ per cent of X-ray sources will have real optical counterparts with $R\lesssim25$, (Georgakakis et al., in preparation), and thus could not be identified as LBGs and included in our sample due to a chance alignment. The above estimates should therefore be corrected by a factor $\sim 0.3$, and we in fact expect $\sim 1.0$ false X-ray detected LBGs in our entire survey.

The samples of detected objects were then compared to previously identified objects in the fields. All but one of the X-ray detected LBGs in our sample from the \hdfn\ and \gws\ have been optically identified previously, and have spectroscopic redshifts \citep{Laird06,Steidel03}. One object has also been optically identified by \citet{Stern02} in the \lynx\ field . When no spectroscopic confirmation is available, we assign a source a redshift of $z=3$, approximately corresponding to the peak of the optical selection function (see section \ref{sec:selfunc}). 
Each spectroscopically identified source is optically classified as a galaxy (GAL: no optical AGN signature), narrow-line AGN (NLAGN: strong Ly$\alpha$, significant C\textsc{IV} $\lambda 1549$ emission lines, no emission line with FWHM$>$2000 km s$^{-1}$), or broad-line QSO (any emission line with FWHM$>$2000 km s$^{-1}$).
 
To perform the X-ray photometry we use the counts within the 95 per cent radius, as the additional counts improve the statistics. These counts are background subtracted and divided by the average value of the exposure map within the extraction region, to calculate a count rate corrected for instrumental effects (e.g. varying exposure times, mirror vignetting, quantum efficiency). This count rate was converted to a flux in the same band, and rest-frame $2-10$ keV X-ray luminosities are calculated using the assigned redshift, assuming a power-law with $\Gamma=1.9$ and the appropriate Galactic $\mathrm{N_H}$ (see Table \ref{tab:fields}).
$2-10$ keV rest-frame energies are well matched to the observed $0.5-2$ keV energies at \z3, thus large, uncertain $k$-corrections are not necessary.
Errors on the fluxes and luminosities are calculated from the Poissonian errors on the total observed counts, calculated according to \citet{Gehrels86}. There is additional uncertainty in the luminosity measurements due to the assumption of an intrinsically unabsorbed power-law, and the assumed redshifts for objects without spectroscopy. These effects are small ($<10$ per cent) for moderate absorbing column densities ($\mathrm{N_H} \lesssim 10^{22} \mathrm{cm^{-2}}$) but could lead to underestimates of the luminosities of more absorbed sources. We discuss the possibility of an absorbed population further in section \ref{sec:discuss}.
We find a total of 22 X-ray detected LBGs. The source properties are given in Table \ref{tab:objects}. 
%
%
\begin{table*}
\caption{
X-ray detected Lyman-break galaxies. Column (1): Object identification \citep[for \gws\ and \hdfn\ adopting designations from][]{Steidel03,Laird06,Reddy06}. Column (2): total magnitude in the \R-band, measured in the Kron aperture (see section \ref{sec:optdet}). Column (3) Net X-ray counts above background in the soft (0.5-2 keV) band, within the 70 per cent PSF radius. Column (4): soft (0.5-2 keV) X-ray flux. Column (5): optical classification. QSO - contains broad emission lines ($>$2000 km s$^{-1}$), AGN - narrow-lined ($<$2000 km s$^{-1}$) AGN spectrum, GAL - no signature of an AGN, UID - unidentified spectroscopically. Column (6): redshift for objects which have been spectroscopically identified. Column (7): 2-10 keV X-ray luminosity, determined from the soft-band flux assuming a power law spectrum with $\Gamma=1.9$, and the given redshift, or $z=3$ for UID sources.}
\begin{tabular}{l c c c c c c c c}
\hline
\multicolumn{1}{c}{Object ID}	& \R\ magnitude	 	& Net counts & Soft X-ray flux		& Optical	& Redshift & X-ray luminosity \\
		&				&   & ($10^{-16}$ \ergscm) 	& type	& 		& ($\log L_\mathrm{X}$(2-10 keV)/\ergs) \\
\multicolumn{1}{c}{(1)}		& (2)		& (3)			& (4)						& (5)		& (6)  & (7)\\
\hline
HDFN-C14    &  	24.92  & 11.81 &$  0.55^{+0.14} _{-0.11} $  &	GAL &   2.981$^a$   & $42.69 ^{+0.10} _{-0.10} $  \\
HDFN-MD39 &   	20.50  & 1888.95 &$ 61.29^{+1.27} _{-1.24} $  &	QSO &   2.583$^b$   & $44.60 ^{+0.01} _{-0.01} $  \\
HDFN-D21$^*$  & 25.37  & 3.78 &$  0.14^{+0.08} _{-0.06} $  &	UID  &   \ldots & $42.10 ^{+0.20} _{-0.22} $  \\
HDFN-M9$^*$   & 24.45  & 7.00 &$  0.27^{+0.13} _{-0.09} $  &	GAL &   2.975$^a$   & $42.38 ^{+0.17} _{-0.18} $  \\
GWS-M47    & 	24.93  & 8.87 &$  3.40^{+1.08} _{-0.84} $  &	AGN &   3.026$^a$   & $43.50 ^{+0.12} _{-0.12} $  \\
GWS-C50    & 	23.87  & 5.70 &$  2.30^{+0.88} _{-0.65} $  &	GAL &   2.910$^a$   & $43.29 ^{+0.14} _{-0.15} $  \\
GWS-D54    & 	22.84  & 23.34 &$  6.83^{+1.34} _{-1.13} $  &	QSO &   3.199$^a$   & $43.86 ^{+0.08} _{-0.08} $  \\
GWS-oMD13  &   	23.46  & 13.87 &$  4.06^{+1.16} _{-0.92} $  &	QSO &   2.914$^a$   & $43.54 ^{+0.11} _{-0.11} $  \\
GWS-MD106  &   	22.66  & 49.83 &$ 15.68^{+2.06} _{-1.83} $  &	QSO &   2.754$^a$   & $44.07 ^{+0.05} _{-0.05} $  \\
Lynx-1     &	20.95  & 68.84 &$ 26.76^{+2.85} _{-2.58} $  &	QSO &   3.093$^c$   & $44.42 ^{+0.04} _{-0.04} $  \\
EGS1-1     &	23.51  & 22.87 &$ 13.91^{+2.73} _{-2.31} $  &	UID &   \ldots & $44.10 ^{+0.08} _{-0.08} $  \\
EGS1-2     &	23.94  & 8.85 &$  2.77^{+1.12} _{-0.82} $  &	UID &   \ldots & $43.40 ^{+0.15} _{-0.15} $  \\
EGS1-3     &	24.09  & 18.31 &$  6.98^{+1.74} _{-1.41} $  &	UID &   \ldots & $43.80 ^{+0.10} _{-0.10} $  \\
EGS1-4     &	20.37  & 70.94 &$ 33.70^{+3.62} _{-3.28} $  &	UID &   \ldots & $44.49 ^{+0.04} _{-0.04} $  \\
EGS1-5     &	24.15  & 7.64 &$  3.88^{+1.34} _{-1.02} $  &	UID &   \ldots & $43.55 ^{+0.13} _{-0.13} $  \\
EGS1-6     &	24.36  & 10.84 &$ 15.88^{+4.05} _{-3.28} $  &	UID &   \ldots & $44.16 ^{+0.10} _{-0.10} $  \\
EGS1-7     &	24.27  & 16.21 &$  9.72^{+1.88} _{-1.59} $  &	UID &   \ldots & $43.95 ^{+0.08} _{-0.08} $  \\
EGS1-8     &	23.79  & 20.01 &$ 11.00^{+2.12} _{-1.80} $  &	UID &   \ldots & $44.00 ^{+0.08} _{-0.08} $  \\
EGS1-9     &	23.22  & 196.46 &$108.09^{+6.07} _{-5.75} $  &	UID &   \ldots & $44.99 ^{+0.02} _{-0.02} $  \\
LALA-1     &	22.38  & 26.50 &$ 13.87^{+2.42} _{-2.08} $  &	UID &   \ldots & $44.10 ^{+0.07} _{-0.07} $  \\
LALA-2     &	23.75  & 7.78 &$  2.71^{+1.16} _{-0.84} $  &	UID &   \ldots & $43.39 ^{+0.15} _{-0.16} $  \\
LALA-3	&	24.27  & 4.66 &$  3.35^{+1.28}_{-0.95} $   & 	UID &   \ldots & $43.49 ^{+0.14}_{-0.15}$\\

\hline
\multicolumn{6}{l}{$^*$ Suspected to be starburst galaxies from the X-ray to optical flux ratio (see section \ref{sec:starburst}).}\\
\multicolumn{6}{l}{$^a$ Redshift given by \citet{Steidel03}, $^b$ \citet{Laird06}, $^c$ \citet{Stern02}.}\\
\end{tabular}
\label{tab:objects}
\end{table*}
\subsection{High luminosity sample}
\label{sec:brightsample}
Our Lyman-break selected sample of AGN contains very few bright objects ($L_\mathrm{X}>10^{45}$\ergs), due to the comparatively small area covered, and we are therefore unable to constrain our fit to the XLF at bright luminosities. To remedy this, we include bright objects ($f_\mathrm{X}>5\times 10^{-15}$\ergscm) from the various large area, X-ray selected surveys described by \citet{Hasinger05}. The CDF-N sample is excluded to avoid overlap with our own analysis. The X-ray detection in these samples was performed in the soft ($0.5-2$ keV) band, and is thus consistent with our work. Full details of the X-ray data analysis are given by \citet{Hasinger05} and references therein.

Spectroscopic identifications in these surveys and at these bright fluxes are highly complete ($\gtrsim 90$ per cent). We include all objects from the parent samples with redshifts $2.5<z<3.5$, whereas the \citet{Hasinger05} luminosity functions were constructed using only unabsorbed `Type 1' AGN. In reality, at these bright fluxes and redshifts all the AGN in the sample are classified as Type 1. 
This may be due to luminosity dependence of the absorbed fraction \citep[e.g.][]{Steffen03}, but the remaining unidentified X-ray sources could correspond to a population of high redshift Type 2 AGN which will be missing from our bright sample. We discuss this issue further in section \ref{sec:discuss}.
Our high luminosity sample consists of 7 objects with $L_\mathrm{X}=10^{44.5-45}$\ergs, and 5 which are brighter than $L_X=10^{45}$\ergs.
\section{Selection functions}
\label{sec:selfunc}
\subsection{X-ray selection function}
The cosmological volume of our survey will depend on the depth of our X-ray observations. The sensitivity of \chandra\ data varies dramatically over the field of view, due to a combination of vignetting and degradation of the PSF. To calculate the X-ray selection function we determine the number of counts required within a detection cell to satisfy the detection criterion (false probability $>10^{-4}$) for each pixel of the X-ray image which is also covered by our optical observations. The radius for the detection cell (70 per cent EEF) and the background are calculated using the same method as for the source detection. The count threshold is then converted to a minimum flux for detection. Thus we can calculate the area of our observations sensitive to a source above a given X-ray flux. 

\begin{figure}
\begin{center}
\includegraphics{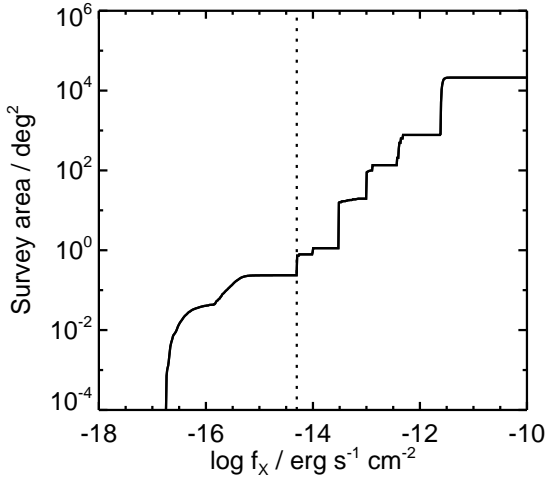}
\end{center}
\caption{
Total area covered by our survey as a function of limiting flux. Above $5 \times 10^{-15}$ \ergscm\ (\textit{dotted line}) we include the surveys used by \citet{Hasinger05}, which dominate the area covered for these brighter fluxes. 
}
\label{fig:acurve}
\end{figure}

For our high luminosity sample we adopt the area curve given by \citet{Hasinger05}, truncated at our flux limit of $5 \times 10^{-15}$\ergs. The X-ray detection was performed using a variety of methods in the different surveys, and include highly significant sources only. However detection is performed in the same energy range ($0.5-2$ keV), and the varying sensitivity limits of these selections are well described by the area curve. Thus by incorporating the sensitivity as a function of flux into our maximum-likelihood fitting (section \ref{sec:MLfit}) we can correct for the effect of incompleteness in the X-ray selection on the survey volume, for the combination of our \chandra\ fields and the various large area surveys in a consistent manner. Figure \ref{fig:acurve} shows the total area curve as a function of soft-band X-ray flux.

%

\subsection{Optical selection function}
\label{sec:optsel}
Selecting objects using the Lyman-break technique will always miss some fraction of objects at a given magnitude due to their true colours not satisfying the selection criteria (colour incompleteness). Photometric errors will also reduce the probability of an object being detected at fainter magnitudes, in addition to scattering objects in or out of the selection box. Determining the optical selection function allows us to correct for these incompletenesses. \citet{Nandra05b} used the effective survey volumes determined by \citet{Steidel99}, appropriate for the selection function for typical starforming LBGs (i.e. GAL classification above). In reality, approximately two thirds of the X-ray sources have AGN or QSO classifications \citep{Laird06}, and these will have different distributions of intrinsic colours, and hence different selection functions. In the present work we have calculated the selection functions for each class independently. The efficiency of the selection function is also highly dependent on the depth and quality of the optical data. The \citet{Steidel99} volumes were averaged for their set of fields.
We determine selection functions appropriate for each individual field in our survey.

We have calculated the selection functions using Monte Carlo simulations \citep{Steidel99,Hunt04}. The first step was to determine an intrinsic distribution of expected colours. GAL colours were generated using a \citet{Bruzual03} model template spectrum for a galaxy with continuous star formation and age 1 Gyr. This template spectrum was reddened using the \citet{Calzetti00} relation for obscuration by dust, with various extinction coefficients drawn from the distribution given in \citet{Adelberger00} which is based on the observed range of LBG spectral shapes. The templates were then redshifted to distribute evenly over the range $z=2-4$. Intergalactic absorption was included using the \citet{Madau95} extinction curve. Finally the template spectra were multiplied by our filter transmissivities to generate the model colour distribution.

\begin{figure}
\begin{center}
\includegraphics{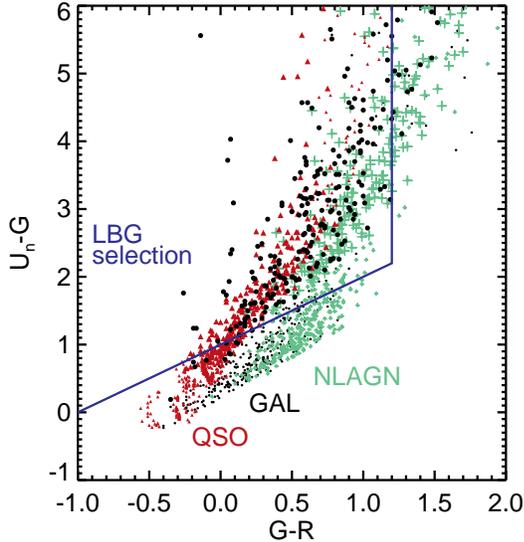}
\end{center}
\caption{
Model colour distributions for different optical classifications: QSO ({\it triangles}), NLAGN ({\it crosses}), and GAL ({\it circles}). Colours are simulated for redshifts $2.0\le z \le 4.0$ (see section \ref{sec:selfunc}). Larger symbols denote objects with $2.7\le z \le 3.3$. While many of the colours falling outside the Lyman-break selection box (\textit{solid line}) correspond to objects at $z\lesssim 2.5$ or $z\gtrsim 3.5$, there is additional scatter due to the range of intrinsic spectral shapes. 5000 model colours are determined for each type (a sample is shown in the figure). These colour distributions are used to add simulated objects to the optical data, which introduces additional scatter due to photometric errors and blending of sources.
}
\label{fig:coldist}
\end{figure}

The colours of NLAGN in LBG samples are generally found to be redder than the typical LBG \citep{Steidel02}. We model the expected range of spectral shapes using the same approach as for galaxies, but drawing from a Gaussian distribution of extinction coefficients, roughly based on the observed range of UV spectral slopes for Lyman-break selected narrow-line AGN \citep{Steidel02,Steidel03}. Such modelling is not a full physical description, being based only on a template spectrum for star-forming processes, and does not explicitly include any contribution due to the AGN. However, it is sufficient to simulate the observed range of spectral shapes and redder distribution of broadband colours for NLAGN. For QSOs we have followed the work of \citet{Hunt04}. The template spectra were generated by varying the continuum slope and Ly$\alpha$ equivalent widths of a composite of 59 QSO spectra from \citet{Sargent89}, based on the Gaussian distributions given in \citet{Hunt04}. Our model colour distributions are shown in Figure \ref{fig:coldist}. 

Simulations were then performed for each optical classification, in which a colour was drawn at random from the appropriate distribution, and used to add artificial objects to the images for each field. Objects were simulated as point sources with seeing appropriate for the field, and \R-band magnitudes drawn from the range 18.5$\leq$\R$\leq$26. This allows photometric errors in the \R-band to scatter objects into the selection range. The objects were recovered using the same detection method used for the original data (section \ref{sec:optdata}). 500 objects were added to each image, and the process was repeated 20 times. The selection function, $P(z,{\cal R})$, was calculated for a number of magnitude and redshift bins as the fraction of input objects which are detected and have colours satisfying the LBG criteria. 
\begin{figure}
\begin{center}
\includegraphics{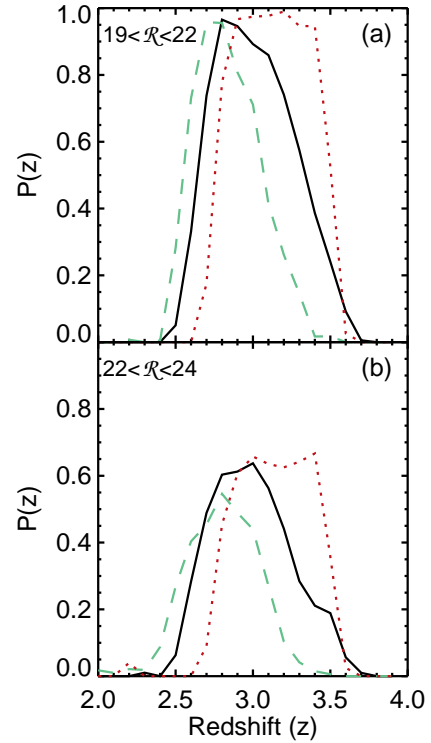}
\end{center}
\caption{
Optical selection functions using the Lyman-break technique, shown here for the Lynx field optical data. The redder colours of NLAGN ({\it dashed line}) result in a less efficient selection function shifted to lower redshifts than for galaxies ({\it solid line}). The lack of an intrinsic Lyman-break in the spectrum of QSOs moves the selection function to higher redshifts ({\it dotted line}). The colours of QSOs are distributed over a smaller range than for galaxies or NLAGN, leading to a more sharply defined selection function. For bright objects (a), the main source of incompleteness is the colour selection. At fainter magnitudes (b), the selection functions are less efficient due to detection incompleteness and larger photometric errors, and is dependent on the depth and quality of the optical data for a field.
}
\label{fig:optsel}
\end{figure}

The selection functions at bright magnitudes for the Lynx field optical data are shown in Figure \ref{fig:optsel} (a). 
Our survey varies in efficiency depending on redshift and the optical type of the AGN host.
The GAL selection function has a mean redshift of $z\approx 3.0$ and samples $\sim 65$ per cent of the total available co-moving volume for a top-hat function for $z=2.5-3.5$.  Due to their intrinsically redder colours, the NLAGN selection function is less efficient, sampling $\sim 50$ per cent of the available volume, and has a lower mean at $z \approx 2.8$. A strong Lyman-break is not present in the intrinsic spectra of QSOs, but is introduced due to intergalactic absorption. This results in a higher redshift selection function, with a mean redshift $z \approx 3.1$. The colours are also distributed over a smaller range, resulting in a more sharply defined selection function, sampling $\sim 70$ per cent of the $z=2.5-3.5$ volume, but almost 100 per cent for $z=2.9-3.4$. 
Thus our survey will select a higher proportion of the overall population of QSOs, biased towards slightly higher redshifts, compared to NLAGN. Determining the selection functions allows us to account and correct for these varying efficiencies.

At brighter magnitudes (${\cal R} \lesssim 22$) photometric errors are small, and the selection functions vary little between fields. At fainter magnitudes, all the selection functions are reduced in efficiency and widened, accounting for the larger photometric errors increasing scatter, and the increased difficulty of detecting faint objects (see Figure \ref{fig:optsel}b). This is highly dependent on the quality of the optical data in each bandpass. For our highest quality data (\hdfn), at \R$\approx 24.0-24.5$ the selection function samples $\sim 75$ per cent of the volume sampled at bright magnitudes, and $\sim 55$ per cent at \R$\approx 25.0-25.5$. Thus  we still sample $\gtrsim 30$ per cent of the total $z=2.5-3.5$ volume. In contrast, in the \egs\ field, with the poorest quality photometry, the efficiency is $\sim 50$ per cent that of the bright selection function at \R$\approx 24.0-24.5$, and $\sim 20$ per cent at \R$\approx24.5-25.0$, for the faintest objects we select in this field. Our incompleteness corrections are therefore large, but thanks to our simulations are well defined over our range of optical magnitudes.

Our high luminosity sample is not subjected to a colour selection criteria. Instead we rely on spectroscopic follow-up of all X-ray sources in the survey fields to have identified a complete sample of objects with $2.5<z<3.5$. Thus, we assume the `optical selection' for this sample is a top-hat function over $2.5<z<3.5$ which selects all X-ray sources in the redshift range. This is a reasonable assumption given the very high completeness of the samples at bright X-ray fluxes. This can then incorporated into our analysis in an analogous manner to the incomplete selection functions for the Lyman-break sample.

\section{Luminosity function}
\label{sec:lumfunc}

A great variety of methods have been used in the past to calculate luminosity functions. One of the simplest and most widely used is the $1/V_{\mathrm{max}}$ method \citep{Schmidt68}. Each object is assigned a maximal survey volume corresponding to the redshift interval over which the source could be included in the sample, given a redshift selection range and the flux limits of the survey. Usually binned luminosity functions are constructed by summing the contribution of all objects within a luminosity bin:
\begin{equation*}
\Phi=\sum_{\Delta \log L_\mathrm{X}} \frac{1}{V_{\mathrm{max}}}
\end{equation*}
However, significant errors are introduced by this method when the objects are close to the flux limits \citep{Page00}, and it is poorly suited for low numbers of sources. Our sample is small, and the flux limits of our survey vary widely between different fields, as well as over a single field of view. We therefore take a different approach, using a maximum likelihood fitting method. No binning is required, so information on individual sources is retained. We are also able to include completeness corrections, correcting for the varying sensitivities of both the X-ray and optical data. In section \ref{sec:bin} we calculate improved binned estimates of the luminosity function using the $N_\mathrm{obs}/N_\mathrm{mdl}$ method \citep{Miyaji01,Hasinger05} for display purposes and comparison with other work.

\subsection{Maximum likelihood fitting}
\label{sec:MLfit}

To perform a maximum likelihood fit we use the STY \citep{Sandage79} estimator, which requires an assumed functional form for the luminosity function. We adopt a broken power-law for the differential luminosity function, which has been found to describe the shape well at lower redshifts \citep[e.g.][]{Barger05}.
\begin{equation}
\phi(L_\mathrm{X})= \dfrac{\dd \Phi(L_\mathrm{X})}{\dd \log L_\mathrm{X}} \propto \left[  \left(\dfrac{L_\mathrm{X}}{L_*} \right)^{\gamma_1} + \left(\dfrac{L_\mathrm{X}}{L_*}\right)^{\gamma_2} \right]^{-1}
\label{eq:lf}
\end{equation}
where $L_\mathrm{X}$ is the 2-10 keV (rest-frame) X-ray luminosity.
The probability of observing a source of a particular luminosity, $L_\mathrm{X}$, is then given by
\begin{equation}
p_i= \frac{ \phi(L_i)}
           { \int_{42}^{48} \int_{2.5}^{3.5} \phi(L_\mathrm{X})\; A(L_\mathrm{X},z)\; C(L_\mathrm{X},z) \dfrac{\dd V}{\dd z} \; \dd z \; \dd \log L_\mathrm{X} }
\label{eq:mlprob}
\end{equation}
The denominator is the expected number of sources at \z3, found by integrating $\phi(L_\mathrm{X})$ over the survey volume. $A(L_\mathrm{X},z)$ is the total area from all our fields sensitive to the flux from a source of luminosity $L_\mathrm{X}$ at redshift $z$. The luminosity to flux conversion assumes an unabsorbed power-law with $\Gamma=1.9$ and the appropriate Galactic $\mathrm{N_H}$ for each field, and thus is consistent with our original calculation of luminosities from observed flux. $\frac{\dd V}{\dd z} {\dd z}$ is the differential co-moving volume element per unit area.

\begin{figure}
\includegraphics{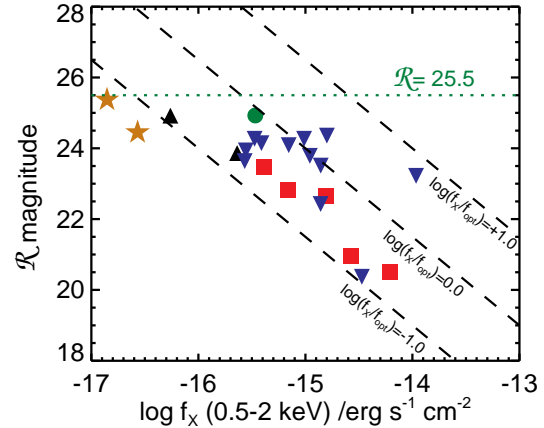}
\caption{
X-ray flux (0.5-2 keV) vs. \R\ magnitude for the sample of X-ray detected LBGs. \textit{Squares}: QSO optical classification; \textit{circles}: NLAGN; \textit{triangles}: GAL; \textit{inverted triangles}: no optical identification. The majority of our sources are found to be distributed between $-1.0<\log (f_\mathrm{X}/f_{\mathrm{opt}})<+1.0$, the expected range associated with AGN. The Lyman-break selection limits the faintest \R-magnitude at which we can detect a source to \R$<25.5$ (\textit{dotted line}), necessitating the incompleteness correction given by equation \ref{eq:fxfoptcorr}. The two faintest objects in our sample (\textit{stars}) are suspected to be star-burst galaxies \citep[][section \ref{sec:starburst}]{Laird06}.
}
\label{fig:fxR}
\end{figure}

Two further corrections for incompleteness effects are included in our calculations. In Equation \ref{eq:mlprob} the factor $C(L_\mathrm{X},z)$ reduces the expected number of sources, accounting for objects fainter than the optical magnitude limit. Figure \ref{fig:fxR}, which shows the optical \R\ magnitude vs. soft band flux illustrates the need for this correction. Lines of constant X-ray to optical flux ratios are plotted, according to the relation given in \citet{Hornschemeier01}, converted for \R\ magnitudes using the relation in \citet{Steidel93} (with no optical colour correction term).
\begin{equation}
\log \left(\frac{f_\mathrm{X}}{f_{\mathrm{opt}}}\right) = \log f_\mathrm{X} + 5.4 + \frac{\cal R}{2.5}
\label{eq:fxR}
\end{equation}
For AGN, this ratio is distributed over the range $-1.0<\log (f_\mathrm{X}/f_{\mathrm{opt}})<+1.0$ over several orders of magnitude in flux \citep[e.g.][]{Schmidt98,Akiyama00,Alexander01}. Our sample of X-ray detected LBGs fall within this range. However, at faint X-ray fluxes ($f_\mathrm{X}\lesssim 10^{-15}$\ergscm) we would expect some fraction of the sources to have optical magnitudes \R$>25.5$, and so would not be found using our Lyman-break selection. The $C(L_\mathrm{X},z)$ factor in equation \ref{eq:mlprob} attempts to correct for this, by reducing the expected number of sources in proportion to the fraction of the $-1.0<\log (f_\mathrm{X}/f_{\mathrm{opt}})<+1.0$ falling below our optical magnitude limit,
and assuming an even distribution over this range. Some small fraction of extreme AGN may reside outside this range \citep[e.g.][]{Hornschemeier01,Barger03c,Lehmer05b}, but the true distribution is unknown, necessitating this simplifying assumption.
Thus we adopt,
\begin{equation}
C(L_\mathrm{X},z)=\frac{{\cal R}_{\mathrm{lim}}}{5} + \frac{1}{2}(1.0+\log f_\mathrm{X}(L_\mathrm{X},z) +5.4) 
\label{eq:fxfoptcorr}
\end{equation}
This correction is directly related to the X-ray flux, and thus is defined for every value of $L_\mathrm{X}$ and $z$. 
For fluxes $\sim 10^{-15}$\ergscm, corresponding to the median flux in our sample, we expect 20 per cent of the sources to have ${\cal R}>25.5$, so this correction is comparatively small, but rises for the fainter fluxes probed by our survey. At the limit of our $\sim 200$ ks fields, $f_\mathrm{X} \sim 2 \times 10^{-16}$\ergscm,  the correction is a factor $\sim 2$. At the extremely faint X-ray fluxes probed in the \hdfn\ we actually expect all the optical counterparts of AGN to be fainter than our limit. Thus, we suspect the 2 faintest objects in the \hdfn\ have a starburst origin, which we discuss further below, and exclude them from our fitting of the XLF of AGN.

\begin{figure}
\includegraphics{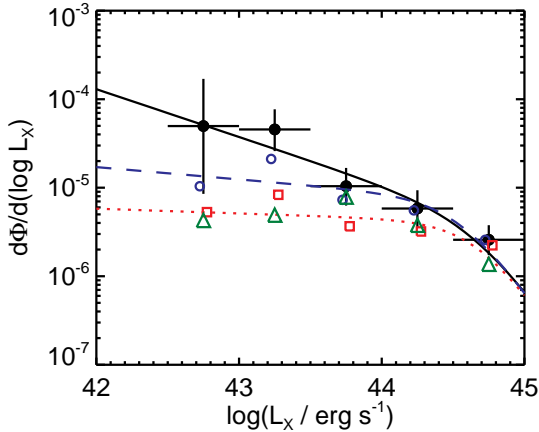}
\caption{
The effect of the completeness corrections on the XLF. \textit{Solid black line/circles}: our best fit XLF and binned estimates, with all completeness corrections applied; \textit{dashed blue line/open circles}: without accounting for objects fainter than ${\cal R}=25.5$ based on the $f_\mathrm{X}/f_{\mathrm{opt}}$ distribution; \textit{dotted red line/open squares}: no correction for optical incompleteness (a top-hat function over $z=2.5-3.5$ is assumed for all optical magnitudes, rather than the Lyman-break selection functions, and there is no correction for objects fainter than ${\cal R}=25.5$). Poissonian error bars are only shown for the best fit, but will be identical for the other points at the same luminosity. Correcting for the incompleteness of the Lyman-break selection increases our determination of the XLF at the faint-end, but does not steepen the slope. Accounting for objects fainter than ${\cal R}=25.5$ results in a steeper faint-end slope. \textit{Open green triangles:} reproduction of the \citet{Barger05} work/method for $z=2.5-3.5$, using only (incomplete) spectroscopic identifications.
}
\label{fig:lfcomp}
\end{figure}

Our second correction accounts for incompleteness within the Lyman-break samples (i.e. at \R$<25.5$). The optical selection functions calculated in section \ref{sec:selfunc}, even at bright magnitudes, are considerably less efficient than a $z=2.5-3.5$ top-hat function, implicitly assumed in the integration limits in Equation \ref{eq:mlprob}. However, this correction depends on the optical classification and \R-magnitude, which is specific to each detected source. Thus, we correct for this effect by introducing an incompleteness weighting, $w_i$, for each object, following \citet{Zucca94}. $w_i$ is inversely proportional to the probability of object $i$ being included in our sample, and thus can be calculated from the ratio of the total survey volume (given the X-ray luminosity of the source, $L_i$), and the effective volume sampled by the selection function.
\begin{equation}
w_i =\frac {V_{\mathrm{tot}}} {V_{\mathrm{eff}}} =
	\frac  { \int_{2.5}^{3.5} A(L_i,z) \dfrac{\dd V}{\dd z} \; \dd z }
		{ \sum_{\mathrm{fields}}\int A_f(L_i,z) P_f({\cal R}_i,z) \dfrac{\dd V}{\dd z}\; \dd z }
\label{eq:wi}
\end{equation}
where  
$P_f({\cal R}_i,z)$ is the selection function for a field at the objects optical magnitude, ${\cal R}_i$, for the optical classification type, as determined in section \ref{sec:selfunc}, and $A_f(L_i,z)$ is the area of that field sensitive to the source luminosity, $L_i$, at $z$. We sum the effective volumes from  each of our fields.
For spectroscopically unidentified objects we must use an average of the three selection functions, weighted according to the fraction of each spectroscopic type expected. \citet{Steidel03} find roughly equal numbers of LBGs classified as NLAGN or QSO in their very large sample of spectroscopically identified objects. \citet{Laird06} also found roughly one third of X-ray detected LBGs exhibited no optical signature of an AGN. Thus we take the average of the three selection functions, with each type weighted equally, for unidentified objects. 
For our high luminosity sample $P_f({\cal R}_i,z)$ is a top-hat function, and thus each object is assigned a weighting of $w_i=1$. Figure \ref{fig:lfcomp} demonstrates the effects of our completeness corrections on the determination of the faint-end slope of the XLF.

The best fit values for $\gamma_1$, $\gamma_2$ and $L_*$ in equation \ref{eq:lf} can then be found by minimizing the log-likelihood function,
\begin{equation}
{\cal S} = -2 \ln{\cal L} = -2 \sum \frac{w_i}{<w>} \ln p_i
\label{eq:mlS}
\end{equation}
Introducing the weights artificially reduces the size of errors estimated from the likelihood fitting, and is therefore balanced by the average weight \citep{Ilbert05}. This does not effect the minimization of $\cal S$ to determine the best fit parameters. 

The overall normalisation of the luminosity function is not determined from the minimization of equation \ref{eq:mlS}, and so is calculated using the best fit model, $\phi_{\mathrm{mdl}}(L_\mathrm{X})$, to match the observed number of sources,
\begin{equation}
K_{\mathrm{norm}}=\frac{ \sum w_i } { \int \int_{2.5}^{3.5} \phi_{\mathrm{mdl}}(L_\mathrm{X})\; A(L_\mathrm{X},z)\; C(L_\mathrm{X},z) \dfrac{\dd V}{\dd z} \; \dd z \; \dd \log L_\mathrm{X} }
\end{equation}
where $\sum w_i$ gives the effective total number of sources, corrected for incompleteness in the optical selection.

We estimate 1$\sigma$ errors on the model parameters from the maximum projection of the $\Delta S = 1.0$ likelihood surface for each parameter. The error in $K_{\mathrm{norm}}$ is estimated from the Poissonian error given the total number of sources. The best fit parameters are given in Table \ref{tab:bfpar}
\begin{table}
\caption{Best fit parameters for the XLF at \z3 derived from the maximum likelihood fitting (section \ref{sec:MLfit}).}
\begin{tabular} {l l}
\hline
Parameter & Value \\
\hline
$\gamma_1$ &     $0.54^{+0.20}_{-0.25}$\\
$\gamma_2$ &    $2.33^{+0.40}_{-0.30}$\\
$\log L_*$ / \ergs      &      $44.68^{+0.27}_{-0.3}$\\
$K_{\mathrm{norm}}$ / Mpc$^{-3}$ & $4.58^{+1.27}_{-1.01} \times 10^{-6}$\\ 
\hline
\end{tabular}
\label{tab:bfpar}
\end{table}
\subsection{Binned estimates}
\label{sec:bin}

\begin{figure}
\includegraphics{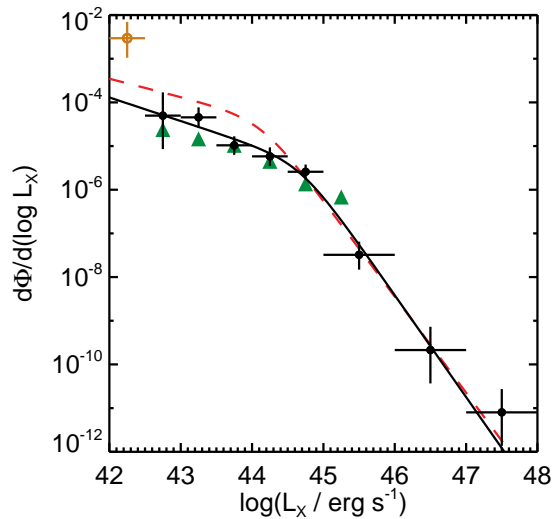}
\caption{
X-ray luminosity function of AGN at \z3. \textit{Black line}: our maximum likelihood fit; \textit{black circles}: binned estimates using the $N_{\mathrm{obs}}/N_{\mathrm{mdl}}$ method; \textit{open orange circle}: $1/V_\mathrm{max}$ estimate for the lowest luminosity bin, containing the suspected star-burst galaxies; \textit{red dashed line}: \citet{Barger05} best fit for a PLE model at $z\le 1.2$, evaluated at $z=1$; \textit{green triangles}: reproduction of the \citet{Barger05} work/method for $z=2.5-3.5$, using spectroscopic and photometric redshifts.
}
\label{fig:lf}
\end{figure}

Binned estimates of the luminosity function are calculated using the $N_{\mathrm{obs}}/N_{\mathrm{mdl}}$ method described by \citet{Miyaji01}. The value within a bin is given by:
\begin{equation}
\phi_{\mathrm{bin}} = \phi_{\mathrm{mdl}}(L_{\mathrm{bin}}) \dfrac{N_{\mathrm{obs}}}{N_{\mathrm{mdl}}}
\end{equation}
where $\phi_{\mathrm{mdl}}$ is the best fit model from the maximum likelihood fitting, evaluated at the bin centre, $L_{\mathrm{bin}}$. $N_{\mathrm{obs}}$ corresponds to the observed number of objects within the bin, but for our analysis this should be corrected for the Lyman-break selection incompleteness. Thus,
\begin{equation*}
N_{\mathrm{obs}}=\sum_{\Delta \log L_\mathrm{X}} w_i
\end{equation*}
$N_{\mathrm{mdl}}$ is the expected number of objects within the bin based on the best fit model.
\begin{equation*}
N_{\mathrm{mdl}}=\underset{\scriptscriptstyle \Delta \log L_\mathrm{X} }{\int} \int_{2.5}^{3.5} \phi_{\mathrm{mdl}}(L_\mathrm{X})\; A(L_\mathrm{X},z)\; C(L_\mathrm{X},z) \dfrac{\dd V}{\dd z} \; \dd z \; \dd \log L_\mathrm{X}
\end{equation*}
This estimator accounts for the varying sensitivity limits within a bin, which are known to result in biased estimates. It offers an improvement over the \citet{Page00} $\phi_{\mathrm{est}}$ method as it allows the luminosity function to vary within a bin. Errors in the binned estimates are calculated from the Poisson uncertainties for the number of objects in each bin. The binned results, along with the best fit, are plotted in Figure \ref{fig:lf}.

%
%

%
%

\subsection{Evolution of the X-ray luminosity function}

\begin{figure}
\includegraphics{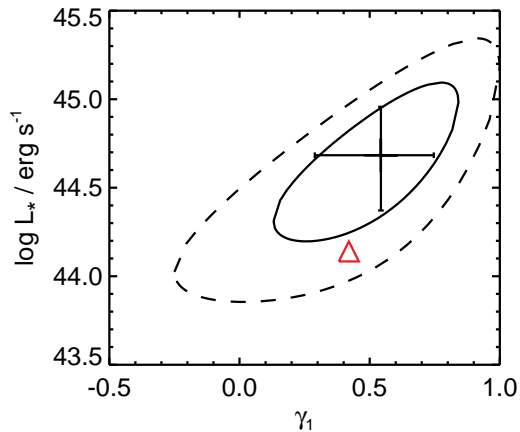}
\caption{
Confidence interval for $L_*$ and $\gamma_1$ for \z3 maximum likelihood fitting ({\it solid}: 68.3  per cent, {\it dashed}: 95 per cent confidence level, error bars indicate 1 dimensional 1 $\sigma$ errors). We also show the \citet{Barger05} PLE model, evaluated at $z=1$ ({\it triangle}).
We find no evidence for a flattening of the faint-end slope (although we cannot rule out the possibility), and weak evidence of continued positive evolution of $L_*$.
}
\label{fig:confplot}
\end{figure}

Our $z=3$ XLF shows excellent agreement with the double power-law model above a luminosity of $\log L_{\rm X} = 42.5$.  Below this we see an upturn, which we comment further upon below. For comparison with lower redshifts, we plot the best fit XLF on Figure \ref{fig:lf} from \citet{Barger05}, evaluated at $z=1$ for a PLE model fit to objects with $z<1.2$. It can be seen that both the bright and faint slopes at $z=3$ are similar to those at $z=1$, but the XLFs are clearly different in that the $z=3$ data points at low luminosity fall below those at $z=1$.
In addition, $L_*$ at $z=3$ appears to be systematically higher. To make a formal comparison, we have determined two-dimensional confidence regions for $\gamma_1$ and $L_*$,   which we show in Figure \ref{fig:confplot}, compared to the \citet{Barger05} parameters for the PLE model. Due to our small sample size, the errors in our best fit parameters are large and dominate over the errors at $z=1$. None the less, our determination of $\gamma_1$ is in good agreement with the Barger value at lower redshifts. We thus find no evidence for a flattening of the faint-end slope of the XLF between redshift $z=1$ and \z3, although we are unable to exclude the possibility of some amount of flattening. We do, however, find weak evidence for continued positive evolution of $L_*$, but the significance of this result is not high. We find no evidence for evolution in the bright-end, although we note that the \citet{Barger05} $z=1$ XLF and our \z3 results are poorly constrained at $L_\mathrm{X}\gtrsim 10^{44.5}$\ergs.

\begin{figure}
\includegraphics{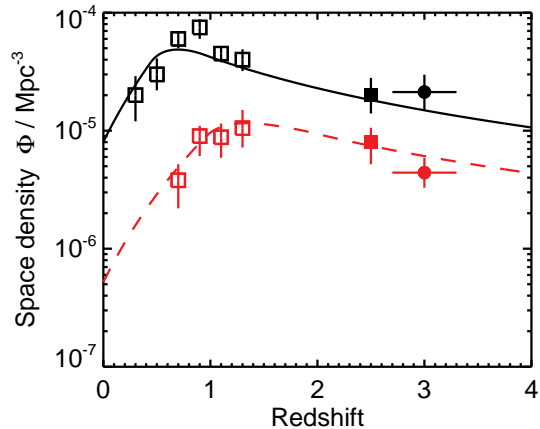}
\caption{
Space density of AGN as a function of redshift, for $L_\mathrm{X}(2-10 \;\mathrm{keV})=10^{43-44}$\ergs (\textit{solid line}) and $L_\mathrm{X}(2-10 \;\mathrm{keV})=10^{44-45}$\ergs (\textit{dashed line}). Lines show the best fit Luminosity Dependent Density Evolution model of \citet{Ueda03}, in which the number density of AGN peaks at \z1, with the peak moving to lower redshifts for lower luminosities. We also show the results from \citet{Barger05} at low redshift (\textit{open squares}), and \citet{Barger05b} for $z=2-3$, using IR and optical photometric redshifts to improve completeness (\textit{solid squares}). Our results (\textit{circles}) using Lyman-break selection are in good agreement with \citet{Barger05b}, and consistent with the decline in number density of lower luminosity AGN above \z1 predicted by LDDE.
}
\label{fig:spacedens}
\end{figure}

Integrating the luminosity functions, we find a total number density of AGN (with $\log L_{\rm X} > 42$) of $1.0^{+0.8}_{-0.4}$ erg s$^{-1}$ Mpc$^{-3}$, a decline by a factor $\sim 3.6$ from $z=1$ to $z=3$. Because of the increase in $L_*$, the decline in the inferred luminosity density is much smaller. We find a total 2-10 keV luminosity density for black hole accretion of $2.2^{+0.5}_{-0.4} \times 10^{39}$ erg s$^{-1}$ Mpc$^{-3}$, about a factor $\sim 2$ smaller than at $z=1$.  In Figure \ref{fig:spacedens} we show results for the space density derived from our best fit XLF at \z3 in two luminosity ranges. There is a clear decline in the space density of these lower luminosity AGN above redshift \z1, and our results are consistent with the \citet{Ueda03} LDDE model.

%
%
\subsection{Pure starburst contribution}
\label{sec:starburst}
The two faintest X-ray sources in our sample, both detected in the \hdfn, have an X-ray to optical flux ratio which places them outside the expected range for AGN. We therefore suspect that the X-ray emission from these objects is due to vigourous starbursts taking place, rather than AGN activity. It is not unexpected that the only detections at these faint fluxes are starbursts, rather than AGN. The \R-band limit for initial pre-selection with the Lyman-break technique means we do probe deep enough with the optical data to sample the AGN $f_\mathrm{X}/f_\mathrm{opt}$ range (see Figure \ref{fig:fxR}). 
One of suspected starbursts (HDFN-M9) was previously reported by \citet{Laird06}. This object has been spectroscopically identified, and is classified as GAL due to the lack of any signature of an AGN in the optical spectrum, which is consistent with a starburst origin for the X-ray emission. 

These objects are contaminants of the AGN sample, and were thus excluded in our fitting of the XLF.
It is however informative to calculate the differential space density in our faintest bin, including these 2 detected objects. The $f_\mathrm{X}/f_{\mathrm{opt}}$ completeness correction (equation \ref{eq:fxfoptcorr}) is not valid for these objects since they are not AGN, and we cannot use the $N_{\mathrm{obs}}/N_{\mathrm{mdl}}$ method to calculate the binned estimate as this is dependent on the fit to the AGN XLF. We therefore use a simpler $1/V_{\mathrm{eff}}$ estimate, where $V_{\mathrm{eff}}$ is calculated as in equation \ref{eq:wi}, and we sum the contribution of the 2 objects for a 0.5 dex luminosity bin. This result is shown in orange in Figure \ref{fig:lf}, and is considerably higher than the AGN XLF. Comparing to the predicted number of AGN detectable in our survey, based on our best fit XLF, we estimate the significance of this excess at $>2\sigma$.

This upturn in the XLF at the faintest luminosities has not previously been seen at \z3, as preceding studies have not probed to sufficiently faint fluxes. Neither of our suspected starburst galaxies were even reported as X-ray sources in the \citet{Alexander03} catalogue of the CDF-N. Given we only detect 2 objects, there is considerable uncertainty in our result. The formal significance of the excess contribution is low, and we are limited to a very small area with sufficiently deep X-ray data. We may also be particularly susceptible to Eddington bias so close to our detection limit. However, detection of an additional suspected starburst galaxy was also reported by \citet{Laird06}, which now falls just below the detection limit when using our updated PSFs and direct extraction of the X-ray counts. 
We thus conclude that at high redshift, such starburst galaxies are likely to be a significant contaminant of the XLF of AGN at the very faintest luminosities.

\begin{figure}
\includegraphics{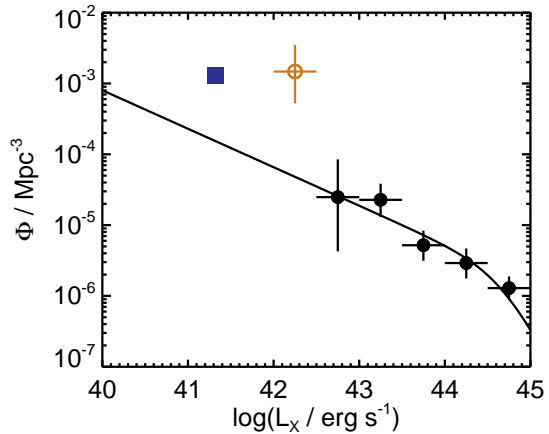}
\caption{
Extrapolation of the faint-end of the integrated luminosity function at \z3 for comparison with the contribution of star-forming galaxies. Both the $1/V_{\mathrm{max}}$ estimate for the number density of the 2 suspected star-burst galaxies (\textit{open circle}) and the estimated number density of star-forming galaxies \citep[\textit{square}, from the stacking analysis of X-ray undetected LBGs by][]{Laird06} fall above our best fit luminosity function.
}
\label{fig:stacking}
\end{figure}

\citet{Laird06} performed a stacking analysis of the X-ray counts from the remaining LBGs in the \hdfn\ without direct X-ray detections, which allows us to probe to yet fainter X-ray luminosities, below our detection limit.
We have derived the space density of these objects, using the $1/V_\mathrm{max}$ method and our GAL selection function at the \R-magnitude of each object. In Figure \ref{fig:stacking} we extrapolate our best fit XLF to lower luminosities, and compare to the space density of the stacked LBGs at their average X-ray luminosity, and our 2 starburst detections. The stacking result also falls above the predicted contribution from AGN. Combined with our direct detections, this provides further evidence for a large population of bright star-forming galaxies at high redshift, which dominate in number density over AGN at these luminosities. This contribution has not been found in lower redshift ($z\lesssim 1.2$) studies \citep[e.g.][]{Barger05} of the XLF, indicating continued evolution of the brightest star-forming galaxies above $z=1$, consistent with findings from UV/optical studies \citep[e.g.][]{Schiminovich05,Reddy07}.

%

\section{Discussion}
\label{sec:discuss}

\subsection{Evolution of AGN at high redshift}
\label{sec:evol}
Our results show that the XLF of AGN continues to evolve above redshifts \z1. This evolution is characterised by a reduction in the number density at faint luminosities ($L_\mathrm{X}<L_*$), although the actual faint-end slope may not flatten. Such evolution cannot be explained as pure luminosity evolution. The faint-end evolution appears more consistent with a reduction in the density. However, there is clearly a luminosity dependence as this evolution is not seen in the bright end. Our result is thus consistent with some form of Luminosity Dependent Density Evolution. However, most of the specific forms of LDDE in the published literature involve a flattening of the faint end slope above \z1 \citep[e.g.][]{Hasinger05, Ueda03}. We find no such evidence, and it is possible that the inference of a flattening seen in previous work is due to incompleteness in the redshift identifications of X-ray sources. 
While our Lyman-break selection also suffers from incompleteness, we have corrected for such effects in our calculations.

The initial results for the space density of moderate luminosity AGN ($L_\mathrm{X}=10^{43-44.5}$\ergs) at \z3 using our method, by \citet{Nandra05b}, showed a markedly higher value than previous estimates, particularly those relying predominantly on spectroscopic redshifts \citep{Cowie03, Barger05}. Our new results are consistent with both those of \citet{Nandra05b} and the revised space density using photometric redshifts by \citet{Barger05b}. While \citet{Nandra05b} found little evidence for a decline in AGN numbers from $z=1$ to $z=3$, the improved analysis and larger sample in our current work strongly indicates that there is such a decline, in support of the conclusions of \citet{Barger05b}.
Indeed, our results should be more robust than either of these previous studies, due both to the improvements in the combined analysis of X-ray sensitivity limits and the optical selection functions, and the much larger survey area and source numbers at faint fluxes. 

We can also compare our results with previous estimates of the XLF at \z3. \citet{Barger05} presented high-redshift XLFs in two bins, $z=1.5-3$ and $z=3-5$, which are not well matched to the redshift range sampled by our Lyman-break selection. We have therefore attempted to reproduce the \citet{Barger05} XLF using the same samples, but for $z=2.5-3.5$. We select soft band detected sources with spectroscopic or photometric redshifts from the CDF-N \citep{Barger03c}, CDF-S \citep{Zheng04}, and \chandra\ Large Area Synoptic X-ray Survey \citep[CLASXS,][]{Steffen04}. The area covered as a function of flux is taken from \citet{Barger05}. The $1/V_{\mathrm{max}}$ method is used to calculate the binned XLF, thus,
\begin{equation}
\Phi_{\mathrm{Barger}}=\sum_{i=0}^{N_{\mathrm{bin}}} \frac{1}{\int^{3.5}_{2.5} A(L_i,z) \dfrac{\dd V}{\dd z} \; \dd z }
\end{equation}
where the summation is over all $N_{\mathrm{bin}}$ sources in the luminosity range for each bin. 

Initially we only include sources with spectroscopic redshifts. The results are shown by the green triangles in Figure \ref{fig:lfcomp}. While at luminosities $\gtrsim 10^{43.5}$\ergs\ there is good agreement with our best determination of the \z3 XLF, at lower luminosities there is only agreement with our results if we do not apply completeness corrections. Our corrected result is considerably higher. Including objects with photometric redshifts increases the numbers at lower luminosities, and there is much better agreement with our XLF (see Figure \ref{fig:lf}), although our result is still systematically higher. Clearly using incomplete samples of low luminosity X-ray sources at these redshifts, without applying completeness corrections, can incorrectly indicate a flattened faint-end slope. Photometric redshifts are necessary to increase the completeness, but at the faintest luminosities there may be residual incompleteness which is difficult to correct for. There remain a number of other uncertainties with photometric redshift determinations, including the errors in the photo-z's, possibility of catastrophic failures, and the high probability of chance superpositions of optical counterparts.

We also note the clear co-dependence between determinations of $\gamma_1$ and $L_*$ in Figure \ref{fig:confplot}. Accurate measurements of the faint-end slope are therefore essential to determine $L_*$ accurately, which is often used as a tracer of the more fundamental underlying black-hole/host galaxy mass distribution \citep[e.g.][]{Hopkins06b}. Thus, careful completeness corrections for the detection of faint AGN, to ensure that the faint-end slope is not artificially flattened, are essential when using the  luminosity function to investigate AGN evolution.

While we find marginal evidence for positive evolution of $L_*$, we find no evidence for evolution of the XLF at brighter luminosities ($L_\mathrm{X}\gtrsim 10^{45}$\ergs), 
which has previously been suggested by a number of authors \citep[e.g.][]{Ueda03, Hasinger05, LaFranca05}. This may be purely due to uncertainty in our bright-end determination, where we have small numbers of sources.
A high luminosity, high redshift population may be missed due to remaining incompletness in the spectroscopic follow-up of our high luminosity sample, although this will also affect previous studies.
There is evidence that the number density of the brightest AGN peaks around $z\approx2$, but declines again at higher redshifts \citep[e.g.][]{Hasinger05,Silverman05,Silverman07b}, in a similar manner to optically selected QSOs \citep[e.g.][]{Wolf03,Croom04b,Bongiorno07}, which could explain the apparent lack of evolution when comparing $z=1$ and $z=3$. 
The lack of observed evolution may indicate that we miss a population of luminous, absorbed AGN (i.e. Type 2 QSOs) which are detected in the hard band at $z=3$. The lack of AGN in our high luminosity sample which exhibit optical signatures of absorption supports this conclusion. Such a population may not be identifiable at $z=1$ as the hard band probes lower rest-frame energies.

\subsection{Limitations of the LBG selection method}

We believe our results using the Lyman-break method to select, and correct for incompleteness in high redshift samples of X-ray detected AGN, are the most robust and accurate yet, but there are limitations to our approach. The necessary completeness corrections are well defined (an advantage over direct identification methods), but they can be quite large, particularly for the faintest X-ray luminosities. At the $L_\mathrm{X}=10^{43-43.5}$\ergs\ our typical overall correction is a factor $\sim 4$, accounting for both the incompleteness in the Lyman-break selection function for the typical \R-magntude, and the expected fraction of sources fainter than our optical limit. This incompleteness means we require a larger area coverage to obtain comparable sample sizes to directly identified samples of high redshift X-ray sources, although the incompleteness of such samples will also be high, and more difficult  to correct.

In section \ref{sec:optsel} we derived selection functions for different optical classifications, and noted the differences in redshift and overall efficiency. However, we lack spectroscopic confirmation for more than half our sample, and thus use an average of our 3 selection functions. Spectroscopy would also allow us to determine the true redshifts of our unidentified sources, which are currently assigned $z=3$. However, while spectroscopy is clearly desirable, the errors in our results are dominated by the Poissonian uncertainties due to the small number of sources, rather than redshift or optical classification.

A number of other authors \citep[e.g.][]{Ueda03, LaFranca05} have attempted to correct for the effect of intrinsic absorption, due to obscuration within the AGN host, on the X-ray luminosity. We have chosen not to do this for a number of reasons. Our work is focussed on \z3, thus the observed soft ($0.5-2$ keV) emission corresponds to a comparatively hard rest-frame energy ($\sim 2-8$ keV), which should be reasonably unaffected by moderate absorbing column densities ($\mathrm{N_H} \lesssim 10^{22} \mathrm{cm}^{-2}$), and so corrections should be small. Many of our sources contain very few counts, thus there is little spectral information and absorption corrections are uncertain. Additionally, if an absorption correction is applied this removes the simple relation between observed flux and luminosity, which would complicate our analysis in section \ref{sec:MLfit}. Performing X-ray detection and photometry at harder observed energies, but with the same analysis, would be possible to reduce absorption effects, but would substantially reduce the sensitivity to faint fluxes. Indeed, only 11 of our soft X-ray detected LBGs are also detected in the hard band to the same significance level.  None the less, it remains a limitation of our study that we are unable to correct for absorption effects without running the risk of introducing greater error.

\subsection{Census of high redshift AGN}
\label{sec:census}
The comparison between our results and previous estimates of the XLF at high redshift provides an indication of the nature of high-redshift AGN, and their host galaxies. The faint-end of our Lyman-break selected XLF is systematically higher than determined from (incomplete) followup of purely X-ray selected samples. 
Thus the Lyman-break method selects a sample of objects which accounts for the entire known population of X-ray selected, moderate luminosity AGN at $z=3$.
To be selected, these AGN must be bright in the rest-frame ultraviolet, and have similar colours to star-forming LBGs.
This is hardly surprising for bright QSOs, whose UV emission will be dominated by the AGN. For fainter objects, even those with spectra, it is difficult to determine the relative contribution to the UV flux from the AGN accretion activity, and the star-formation processes taking place within the host galaxy. \citet{Steidel02} examined a sample of optically identified AGN in LBG samples, and concluded that the host galaxies of observed AGN could be the equivalent of LBGs without active nuclear activity. Using X-ray emission to identify AGN in the LBG samples finds an additional population which do not exhibit optical AGN signatures, and supports the conclusion that \z3 AGN are hosted by galaxies with similar properties to the LBGs.

This is of great interest, as the majority of moderate luminosity AGN at lower redshifts are found in red host galaxies \citep[e.g.][]{Barger03c,Nandra07b}. \citet{Nandra07b} presented a colour-magnitude diagram for a sample of X-ray selected \z1 AGN, which were found to lie mainly in moderate luminosity galaxies ($-20.5<M_B<22.5$) on the red sequence, at the top of the blue cloud and in the valley in between.
90 per cent of the X-ray sources have rest-frame $U-B>0.6$. By contrast, the absolute magnitude limit of the LBG samples is $M_B\sim 22$ at $z=3$, although our $f_\mathrm{X}/f_\mathrm{opt}$ completeness correction does attempt to account for AGN residing in fainter hosts.

We are unable to directly compare the \z1 colours to our sample, as the rest-frame $U$ and $B$ bands correspond to much redder wavelengths at \z3 than covered by our photometry. 
However, \citet{Shapley05} have performed SED fitting for a sample of rest-frame UV colour selected galaxies, utilising ground based photometry from 0.35 to 2.15 \micron, complemented by Spitzer IRAC data. We have derived rest-frame $U-B$ colours for each of these galaxies, based on the best fit star formation histories. The mean $(U-B)_\mathrm{rest}$ colour is 0.52, with standard deviation 0.23. All of the X-ray selected \z1 AGN (excluding QSOs) are redder than this mean, and 52 per cent lie more than 2 standard deviations away, and are thus redder than over 98 per cent of the rest-frame UV selected galaxies.
Assuming the host galaxies of our \z3 X-ray detected LBG sample have similar colours to the \citet{Shapley05} sample, this tentatively suggests a significant evolution of the host galaxy colours from the blue cloud at $z=3$ to predominantly the red sequence and green valley at $z=1$ \citep[see also][]{Silverman07}.

However, this result does not exclude the possibility of a population of moderate luminosity AGN residing in red host galaxies at $z=3$, but they must be under-represented in both Lyman-break and directly identified samples.
Alternative colour selection techniques have identified populations of red galaxies at high redshifts (e.g. $BzK$ passively evolving galaxies, \citealp{Daddi04}; DRGs, \citealp{Franx03}) which have little overlap with rest-frame UV selected samples \citep{Reddy05}. Such galaxies could host faint, moderately obscured AGN. Our correction for objects fainter than the \R-band limit may partially account for such a population, which will be faint in the rest-frame UV, but it is not possible to determine if they are adequately represented. Thus, if there is a significant population of AGN in red galaxies it will under-represented in our work. Direct spectroscopic identification or photometric redshift determinations will also be difficult due to the faintness of such galaxies, and thus they could also correspond to the unidentified fraction of purely X-ray selected samples.

As discussed in section \ref{sec:evol}, hard-band selected samples of high redshift AGN may include an absorbed, high luminosity population which is missing from our soft-band selected work. Absorption will have an even more severe effect on the faint-end of the XLF, as moderately absorbed sources will not be identified in the soft band, and the hard band will not probe deep enough. Thus a significant population of low luminosity absorbed AGN may remain unidentified at high redshift. 
\citet{Daddi07} and \citet{Georgakakis08} have reported populations of faint, hard sources based on stacking analyses, but the key issue of whether they are Compton-thick is still unresolved.

\subsection{Future prospects}
While our survey covers the largest area of deep ($\gtrsim 200$ ks) X-ray data, and probes to the faintest limits, our total sample is still small, and thus uncertainties remain.
Further work, increasing the sample size, is required to improve constraints on the parameters describing the XLF at \z3. The Lyman-break technique is extremely efficient for identifying AGN in large area X-ray surveys, such as those covered by recent $\sim$200 ks \chandra\ surveys \citetext{e.g. E-CDFS, \citealp{Lehmer05b}, AEGIS-X, PI: Nandra; C-COSMOS, PI: Elvis}.
\citet{Francke08} have already applied similar techniques in the E-CDFS to identify moderate luminosity AGN at \z3 and perform a clustering analysis.
Combined with well-defined completeness corrections these deep, large area surveys should vastly improve the constraints on the faint-end ($L_\mathrm{X}\approx 10^{43-44.5}$\ergs) of the XLF.
Improving constraints at $\gtrsim L_*$ requires even larger area coverage of more moderate depth, and thus the efficiency of our Lyman-break method may also be beneficial for such studies.
A larger area of data of depth $\gtrsim 2$ Ms or greater is required to probe to $L_\mathrm{X}\lesssim 10^{42.5}$\ergs\ and confirm our X-ray detection of a starburst population. However, forthcoming deeper X-ray observations of the AEGIS field (A09, PI: Nandra), taking 3 of the current fields to a depth of 800 ks, will vastly increase the area sensitive to $L_\mathrm{X}\gtrsim 10^{42.5}$\ergs, without the issue of starburst contamination, and provide essential data to improve constraints at the faintest fluxes, currently only accessible by the \chandra\ Deep Fields.
However, to really constrain the high-redshift evolution of the XLF requires large samples, selected over a range of redshifts, and probing to low luminosities, requiring the combination of all surveys over the full range of depths. However, to fully utilise the available depth will require carefully derived completeness corrections, which must therefore be an essential feature of any future studies.

\section{Summary}
\label{sec:summary}

We have presented a new determination of the XLF of AGN at $z=3$, by combining Lyman-break selection in the optical with very deep X-ray observations. This combination allows us to use the full depth of the X-ray data, and to correct accurately for incompletenesses both in the X-ray detection and in the optical photometric selection. As a result, our determination of the faint end of the XLF should be both the deepest, and the most accurate yet obtained. Contrary to some previous results, and LDDE models, we find no evidence for a flattening of the faint end slope at high $z$. There is clear evolution of the XLF, however. This can be described in terms of zero or very mild positive luminosity evolution above $L_*$, and negative density evolution below. 

At the very lowest luminosities probed by our survey ($L_\mathrm{X}\lesssim 10^{42.5}$\ergs) we have identified a population of pure starburst galaxies, which are a significant contaminant of our XLF of AGN. Comparing with previous stacking results confirms the existence of such a population, which is not found at \z1, indicating a rapid increase in the number density of the brightest star-forming galaxies to high redshift.

Our results are higher than previous estimates at the faint end, which we attribute to incompleteness in purely X-ray selected work. Our Lyman-break method identifies blue objects which, after correcting for incompleteness, accounts for all know X-ray sources at $z=3$. This tentatively suggests that the AGN host population is evolving from blue star-forming galaxies at $z=3$, to mainly red galaxies at $z=1$.

Future studies, utilising recent and forthcoming large area, deep surveys, will provide further constraints on the evolution of the XLF of AGN at high redshift. We stress that full consideration of selection and completeness effects is needed for a robust determination of the XLF.

\section*{Acknowledgments}
We thank the referee for comments which improved this paper.
We also thank Naveen Reddy for useful discussions regarding optical selection functions, and Takamitsu Miyaji for providing area curves for our bright sample. JA acknowledges financial support from STFC in the form of a Research Studentship, and a grant from the University of London Central Research Fund. AG acknowledges funding from Marie-Curie Fellowship grant MEIF-CT-2005-025108. ESL acknowledges funding from STFC.
This research has made use of data obtained from the \chandra\ Data Archive and software provided by the \chandra\ X-ray Center (CXC).
%
%
\bibliographystyle{mn2e}

\label{lastpage}
\end{document}